%% file: PanGuideMultilevelResearch_2016-08-04.tex
\pdfoutput=1
\documentclass[man]{apa6}

\usepackage{amsmath}
\usepackage{pdfpages}
\usepackage[breaklinks]{hyperref}
\hypersetup{
  pdfborderstyle={/S/U/W 1},
  colorlinks = false,
  linkbordercolor = {1 1 1},
  urlbordercolor = {1 1 1},
  citebordercolor = {1 1 1}
} 
\usepackage[T1]{fontenc}
\usepackage{textcomp}
\usepackage[american]{babel}
\usepackage{txfonts}
\usepackage{setspace} 
\AtBeginDocument{%
 \abovedisplayskip=10pt plus 2pt minus 6pt
 \abovedisplayshortskip=0pt plus 6pt
 \belowdisplayskip=10pt plus 2pt minus 6pt
 \belowdisplayshortskip=10pt plus 2pt minus 6pt
}

\usepackage[natbibapa]{apacite}
\usepackage{url}
\usepackage{array}
\usepackage[low-sup]{subdepth}
\usepackage{booktabs}
\usepackage{dcolumn}
\usepackage{caption}
\usepackage{graphicx}
\usepackage{color}
\usepackage{inconsolata}

\usepackage{listings}
\definecolor{background}{gray}{0.92}
\definecolor{white}{gray}{1}
\definecolor{black}{gray}{0}
\definecolor{comment}{gray}{0.35}
\definecolor{keyword}{RGB}{0,0,0}
\lstdefinestyle{rcode}{
  backgroundcolor=\color{background},
  commentstyle=\color{comment}\small\ttfamily,
  keywordstyle=\color{keyword}\small\ttfamily,
  frame=leftline,
  rulecolor=\color{background},
  frame=trbl,
  framesep=6pt,
  xleftmargin=6pt
}

\newcommand\widebar[1]{%
  \hbox{%
    \vbox{%
      \hrule height 0.6pt%
      \kern0.256ex%
      \hbox{%
        \kern-0.15em%
        \ensuremath{#1}%
        \kern-0.1em%
      }%
    }%
  }%
} 

\DeclareMathSymbol{\beta}{\mathalpha}{lettersA}{12}
\DeclareMathSymbol{\delta}{\mathalpha}{lettersA}{14}
\DeclareMathSymbol{\epsilon}{\mathalpha}{lettersA}{15}
\DeclareMathSymbol{\varepsilon}{\mathalpha}{lettersA}{34}
\DeclareMathSymbol{\theta}{\mathalpha}{lettersA}{18}
\DeclareMathSymbol{\omega}{\mathalpha}{lettersA}{33}
\DeclareMathSymbol{\rho}{\mathalpha}{lettersA}{26}
\DeclareMathSymbol{\upsilon}{\mathalpha}{lettersA}{29}
\DeclareMathSymbol{\tau}{\mathalpha}{lettersA}{28}
\DeclareMathSymbol{\mu}{\mathalpha}{lettersA}{22}
\DeclareMathSymbol{\nu}{\mathalpha}{lettersA}{23}
\DeclareMathSymbol{\psi}{\mathalpha}{lettersA}{32}
\DeclareMathSymbol{\kappa}{\mathalpha}{lettersA}{20}
\DeclareMathSymbol{\lambda}{\mathalpha}{lettersA}{21}
\DeclareMathSymbol{\phi}{\mathalpha}{lettersA}{30}
\DeclareMathSymbol{\pi}{\mathalpha}{lettersA}{25}
\DeclareMathSymbol{\sigma}{\mathalpha}{lettersA}{27}
\DeclareMathSymbol{\alpha}{\mathalpha}{lettersA}{11}
\DeclareMathSymbol{\gamma}{\mathalpha}{lettersA}{13}
\DeclareMathSymbol{\chi}{\mathalpha}{lettersA}{31}
\DeclareMathSymbol{\eta}{\mathalpha}{lettersA}{17}
\DeclareMathSymbol{\zeta}{\mathalpha}{lettersA}{16}

\title{Multiple Imputation of Multilevel Missing Data: An Introduction to the R Package \texttt{pan}}
\shorttitle{Imputation of Multilevel Data}
\affiliation{Leibniz Institute for Science and Mathematics Education, Kiel, Germany}
\author{Simon Grund$^{1,2}$, Oliver L\"udtke$^{1,2}$, and Alexander Robitzsch$^{1,2}$}
\affiliation{$^1$ Leibniz Institute for Science and Mathematics Education, Kiel, Germany; $^2$ Centre for International Student Assessment, Germany}

\authornote{\noindent Correspondence concerning this article may be directed to Simon Grund, Leibniz Institute for Science and Mathematics Education, 24118 Kiel, Germany; (+49)-431-880-5653; grund@ipn.uni-kiel.de.}

\abstract{
The treatment of missing data can be difficult in multilevel research because state-of-the-art procedures such as multiple imputation (MI) may require advanced statistical knowledge or a high degree of familiarity with certain statistical software.
In the missing data literature, \texttt{pan} has been recommended for MI of multilevel data.
In this article, we provide an introduction to MI of multilevel missing data using the R package \texttt{pan}, and we discuss its possibilities and limitations in accommodating typical questions in multilevel research.
In order to make \texttt{pan} more accessible to applied researchers, we make use of the \texttt{mitml} package, which provides a user-friendly interface to the \texttt{pan} package and several tools for managing and analyzing multiply imputed data sets.
We illustrate the use of \texttt{pan} and \texttt{mitml} with two empirical examples that represent common applications of multilevel models, and we discuss how these procedures may be used in conjunction with other software.
}

\begin{document}

\thispagestyle{empty}
\input{arXiv_titlePage.tex}

\newpage
\setcounter{page}{1}

\maketitle
In recent years years, multilevel models have become one of the standard tools for analyzing clustered empirical data.
Such data often occur in organizational and educational psychology and other fields of the social sciences, for example, when employees are nested within work groups, students are nested within school classes, or in longitudinal studies when measurement occasions are nested within persons.
In addition, empirical data are often incomplete, for example, when participants drop out of the study or do not answer all of the items on a questionnaire.
Several authors have advocated the use of modern missing data techniques such as multiple imputation (MI) rather than traditional approaches such as listwise or pairwise deletion \citep[][]{Allison2001,Enders2010,vanBuuren2012,Newman2014,Schafer2002}.
One central requirement of MI is that the imputation model must be at least as general as the model of interest in order to preserve relationships among variables \citep{Enders2010}.
In the case of incomplete multilevel data, it is important that the imputation model takes the multilevel structure into account in order to ensure valid statistical inferences in subsequent multilevel analyses \citep{Black2011, Graham2012, vanBuuren2011a}.

Although MI is gaining popularity among applied researchers, multilevel imputation models are rarely used in practice.
One of the most commonly recommended software solutions for multilevel imputation is the \texttt{pan} package \citep{Schafer2002a,Schafer2014}, which is freely available in the statistical software R (\citealp{RCoreTeam2015}; see also \citealp{Culpepper2011}).
However, the application of \texttt{pan} can be challenging, and its documentation is rather technical, especially for users who are not familiar with R.
For instance, for multilevel missing data, \citet{Graham2012} recommended ``that you obtain a copy of the PAN program (...), and that you find an expert in R who can help you get started'' (p. 137).

The present paper is intended as a gentle introduction to the \texttt{pan} package for MI of multilevel missing data.
We assume that readers have a working knowledge of multilevel models \citep[see][]{Snijders2012a,Raudenbush2002,Hox2010}.
In order to make \texttt{pan} more accessible to applied researchers, we make use of the R package \texttt{mitml}, which provides a user-friendly interface to the \texttt{pan} package and some additional tools for organizing and analyzing multiply imputed data \citep{mitml032}.
The first section of this paper introduces an empirical example that is used for illustrating the application of \texttt{pan} to multilevel data.
In the following section, we briefly describe the main ideas behind \texttt{pan} and MI, and we discuss which features of multilevel models must be considered when conducting MI.
Finally, we use the \texttt{mitml} package to carry out MI for the empirical example.
In that context, we will discuss possibilities for model diagnostics and tests of nonstandard statistical hypotheses (e.g., model constraints, model comparisons).

\section{Multilevel Modeling: An Empirical Example}

\begin{table}[tbp]
  \begin{threeparttable}
    \caption{Pairwise Observed-Data Correlations Among Variables and Amount of Missing Data}
    \label{tab:dataset}
    \renewcommand{\arraystretch}{1.5}
    \setlength{\tabcolsep}{6.0pt}
    \small
  \begin{tabular}{lrrrrrrr} \toprule
      & MA     & RA     & CA     & SES    & DPM    & DPR    & SC     \\ \midrule
  MA  & & $\phantom{-}0.528$ & $\phantom{-}0.530$ & $\phantom{-}0.232$ & $-0.234$ & $-0.238$ & $-0.217$ \\
  RA  & & & $\phantom{-}0.493$ & $\phantom{-}0.299$  & $-0.291$ & $-0.294$ & $-0.327$ \\
  CA  & & & & $\phantom{-}0.240$ & $-0.265$ & $-0.251$ & $-0.221$ \\
  SES & & & & & $-0.154$ & $-0.155$ & $-0.123$ \\
  DPM & & & & & & $\phantom{-}0.782$ & $\phantom{-}0.399$ \\
  DPR & & & & & & & $\phantom{-}0.419$ \\ \midrule
    Missing Data & 19.4\% & 0\%    & 0\%    & 35.0\% & 61.4\% & 21.5\% & 21.7\% \\ \bottomrule
  \end{tabular}
    \begin{tablenotes}[para,flushleft] ~\\[-1.2ex]
      {\small \textit{Note.} MA = math achievement; RA = reading achievement; CA = cognitive ability; SES = socioeconomic status; DPM = disciplinary problems in math class; DPR = disciplinary problems in reading class; SC = school climate.
      }
    \end{tablenotes}
  \end{threeparttable}
\end{table}

\begin{table}[tbp]
  \begin{threeparttable}
    \caption{Frequent Missing Data Patterns}
    \label{tab:patterns}
    \renewcommand{\arraystretch}{1.5}
    \setlength{\tabcolsep}{8.2pt}
    \small
    \begin{tabular}{lcccccccrrr} \toprule
    Pattern & MA & RA & CA & SES & DPM & DPR & SC & Cases \# & Rel. \% & Cum. \% \\ \midrule
    1 & o & o & o & o & x & o & o & 2306 & 26.3\% & 26.3\% \\
    2 & o & o & o & o & o & o & o & 2134 & 24.3\% & 50.6\% \\
    3 & o & o & o & x & x & o & o & 1173 & 13.4\% & 64.0\% \\
    4 & o & o & o & x & o & o & o & 1125 & 12.8\% & 76.9\% \\
    5 & x & o & o & o & x & x & x & 1027 & 11.7\% & 88.6\% \\
    6 & x & o & o & x & x & x & x & 622 & 7.1\% & 95.7\% \\ \bottomrule
    \end{tabular}
    \begin{tablenotes}[para,flushleft] ~\\[-1.2ex]
      {\small \textit{Note.} The patterns displayed here account for $\geq$ 95\% of the sample. o = observed; x = missing; MA = math achievement; RA = reading achievement; CA = cognitive ability; SES = socioeconomic status; DPM = disciplinary problems in math class; DPR = disciplinary problems in reading class; SC = school climate.
      }
    \end{tablenotes}
  \end{threeparttable}
\end{table}

Multilevel models account for dependencies in the data and allow relationships between variables to be estimated at different levels of analysis or effects that may vary across higher-level observational units.
For the purpose of this article, we assume that the multilevel structure consists of persons (e.g., students, employees) nested within groups (e.g., classes, work groups).
If only the regression intercept varies across groups, the model is referred to as a random-intercept model.
For example, \citet{Chen2002} examined the effects of individual characteristics (e.g., psychological strain) and leadership climate on the self-efficacy of U.S. soldiers.
\citet{Kunter2007} investigated the effects of student- and group-level ratings of classroom management on students' interest in mathematics.
If the effects of additional predictor variables vary across groups, the model is referred to as a random-slope or random-coefficients model.
For example, \citet{Hofmann2003} investigated varying effects of leader-member exchange on safety behavior across work teams in the U.S. army.

The example data set used in this article is from the field of educational research and was taken from the German sample of primary school students who participated in the Progress in International Reading Literacy Study \citep[PIRLS;][]{Mullis2003,Bos2005}.
The data set includes test scores in both mathematics and reading achievement, a measure of cognitive ability, a measure of socioeconomic status (SES), students' ratings of the quality of teaching in their math and reading classes (the prevalence of disciplinary problems), and ratings of the general learning environment (school climate).
For the purpose of this article, we considered only students for whom reading achievement and cognitive ability scores were available, which was true for approximately 99.3\% of the sample (8,767 students in 475
classes).
Ratings of disciplinary problems in math classes were missing for half of the sample due to a planned missing data design \citep{Graham2006}.
Table \ref{tab:dataset} provides an overview of the data set, along with the observed correlations and the percentages of missing values among variables.
Some variables contain additional, unplanned missing data.
In such cases, it is useful to examine the missing data patterns that occur in the data set.
This is shown in Table \ref{tab:patterns}.
Approximately 50\% of the sample adhered to the planned missing data design (Patterns 1 and 2).
In another 25\% of the sample (Patterns 3 and 4), SES was additionally missing.
The remaining patterns were more diverse, and data were missing for math achievement scores, disciplinary problems in reading classes, or school climate.
Planned missing data designs are becoming increasingly popular in large-scale observational studies because such designs can reduce the burden that is placed on each individual participant \citep{Graham2006}.
The missing data mechanism is usually ignorable for variables recorded in this manner, thus enabling us to focus on more specific aspects of MI in multilevel research.

\subsection{Example 1: Random-Intercept Model}

Our first model of interest examined the effect of teaching quality in math classes (disciplinary problems; DPM) on students' math achievement scores (MA).
In addition, we included SES in order to control for differences in socioeconomic background between students and classes.
The student-level variables were centered around the group mean, and the group means were included as predictor variables in order to separate within-group from between-group effects \citep[see][]{Enders2007,Raudenbush2002}.
For student $i$ in class $j$,
\begin{equation} 
  \label{eq:ex1} 
  \mathit{MA}_{ij} = \beta_{0} + \beta_{1} (\mathit{DPM}_{ij} - \overline{\mathit{DPM}}_{j}) + \beta_{2} \overline{\mathit{DPM}}_{j} + \beta_{3} (\mathit{SES}_{ij}- \overline{\mathit{SES}_{j}}) + \beta_{4} \overline{\mathit{SES}}_{j} + \upsilon_{0j} + \epsilon_{ij} \;.
\end{equation}
\noindent
Here, the $\beta$ coefficients denote fixed effects, and $\upsilon_{0j}$ and $\epsilon_{ij}$ denote the residuals at the class and student level, respectively.
We refer to the effects of the average DPM and SES of a class as \emph{between-group} effects, whereas \emph{within-group} effects accounts for the students' individual deviations from that average.
For example, $\beta_4$ denotes the effect of a class' average SES on class-level math achievement, whereas $\beta_3$ denotes the effect of students' individual deviations from the class average on their individual math achievement scores.
The student- and class-level residuals are each assumed to follow a normal distribution with zero mean and variances $\textit{Var}(\upsilon_{0j})$, independently and identically across classes, and $\textit{Var}(\epsilon_{ij})$, independently and identically across students.

\subsection{Example 2: Random-Slope Model}

Our second model of interest examined the relationship between students' cognitive ability (CA) and their math achievement scores.
We assumed that the relationship between the two variables would vary across groups (random slope) because some teachers may nurture students' individual strengths and weaknesses, whereas others may strive to ``equalize'' them.
As before, we included SES to control for differences in socioeconomic background.
In line with recent recommendations for analyzing random-slope variation, we centered the variables around the group means \citep{Hofmann1998,Aguinis2013}.
The group means were included as additional predictors in order to ``reintroduce'' the group-level construct into the model.
The model reads
\begin{equation} 
\begin{aligned}
  \label{eq:ex2} 
  \mathit{MA}_{ij} &= \beta_{0} + \beta_{1} (\mathit{CA}_{ij} - \overline{\mathit{CA}}_{j}) + \beta_{2} \overline{\mathit{CA}}_{j} + \beta_3 (\mathit{SES}_{ij} - \overline{\mathit{SES}}_{j}) + \beta_{4} \overline{\mathit{SES}}_{j} \\
 &+ \upsilon_{0j} + \upsilon_{1j} (\mathit{CA}_{ij} - \overline{\mathit{CA}}_{j}) + \epsilon_{ij} \;,
\end{aligned}
\end{equation}
\noindent where $\upsilon_{1j}$ denotes the random effect of cognitive ability on math achievement per class.
The two random effects (intercept and slope) are assumed to follow a multivariate normal distribution, independently and identically across classes, and the remaining notation is the same as above.

\section{Multiple Imputation of Incomplete Multilevel Data}

Missing data could be addressed by restricting the analyses to completely observed cases (listwise deletion).
However, this approach is more likely to suffer from low power and to give biased results (e.g., \citealp{Little2002}; see also \citealp{Newman2014}).
Multiple imputation has become one of the preferred methods for overcoming these problems \citep{Rubin1987,Schafer2002}.
Using MI, a number of replacements for the missing data are drawn from the distribution of the missing values, given the observed data and an imputation model.
The completed data sets are then analyzed separately, and the results are combined across data sets to form final parameter estimates and inferences \citep[see][for details about the general MI procedure]{Enders2010}.

\subsection{General Aspects of MI}

In most applications of MI, the data are assumed to be missing at random (MAR), a notion that was introduced by \citet{Rubin1976} in his well-known classification of missing data mechanisms.
Consider the hypothetical complete data matrix $\mathbf{Y}$ which is decomposed into observed and unobserved portions $\mathbf{Y} = (\mathbf{Y}_\text{obs}, \mathbf{Y}_\text{mis})$.
An indicator matrix $\mathbf{R}$ denotes whether values are observed or missing.
If the missing data are simply a random sample of the hypothetical complete data, that is, $P(\mathbf{R}|\mathbf{Y})=P(\mathbf{R})$, then the data are missing completely at random (MCAR).
One such scenario occurs in planned missing data designs, where missing values are ``assigned'' randomly to each participant.
If the occurrence of missing data depends on the observed data but missing data occur ``at random'' with these taken into account, that is, $P(\mathbf{R}|\mathbf{Y})=P(\mathbf{R}|\mathbf{Y}_\mathrm{obs})$, then the data are missing at random (MAR).
The two missing data mechanisms MCAR and MAR are often called ``ignorable'' because the exact missing data mechanism need not be known in order to perform MI \citep[for a more general discussion of the role of ``ignorability'', see][]{Enders2010}.
If neither condition holds, that is, $P(\mathbf{R}|\mathbf{Y})=P(\mathbf{R}|\mathbf{Y}_\mathrm{obs},\mathbf{Y}_\mathrm{mis})$, then the data are missing not at random (MNAR).
Most software implementations of MI rely on the assumption that the data are MAR.
Performing MI under MNAR is possible but requires making strong assumptions about the missing data mechanism and is most often used for sensitivity analyses \citep[see][]{Carpenter2013}.
In order to enhance the plausibility of the MAR assumption, it has been suggested that auxiliary variables be included in the imputation model.
These variables are related to either the propensity of missing data or the missing values themselves, without necessarily being part of the model of interest \citep{Collins2001}.
In our empirical example, some data are missing by design and are thus MCAR.
For the remaining data, we will assume that the data are MAR, given the observed portions of the data that can be included as auxiliary variables.

Furthermore, the imputation model must be at least as complex as the analysis model.
If variables or parameters that are relevant for the analysis model are not included in the imputation model, then the procedure could yield biased results \citep{Meng1994,Schafer2003}.
For example, assume that a researcher is interested in testing an interaction between two variables in a multiple regression analysis with partially missing data.
In this case, it would be important that the interaction effect (i.e., product term) is incorporated in the imputation model \citep{Enders2014}.
Similarly, if one is interested in estimating the intraclass correlation (i.e., the variance within and between groups) with incomplete data, it would be crucial to take into account the clustered data structure \citep{Taljaard2008}.
If the model of interest includes random slopes, then the imputation model should allow for different slopes across groups.
Choosing an appropriate imputation model can be challenging, and it may be tempting to resort to ad hoc methods for treating multilevel missing data \citep{Graham2012}.
For example, it has been suggested that the multilevel structure be represented by creating a set of dummy indicator variables \citep{Graham2009,White2011,Drechsler2015}.
In this approach, the dummy indicators are included in the single-level imputation model, and a separate intercept (or fixed effect) is estimated for each group.
However, recent simulation research has indicated that such methods can distort parameter estimates and standard errors in multilevel analyses \citep{Andridge2011,Enders2016,Ludtkeinpress}.

Two broad approaches to performing MI can be distinguished.
In the joint modeling approach, a single statistical model is used for imputing all incomplete variables simultaneously \citep[e.g.,][]{Schafer2002a}.
In contrast, in the fully conditional specification of MI, each variable is imputed in turn using a sequence of imputation models \citep{vanBuuren2011}.
In the present article, we focus on the \texttt{pan} package, which follows the joint modeling paradigm \citep[for a discussion, see][]{Carpenter2013}.

\subsection{The Multivariate Linear Mixed-Effects Model}

The statistical model underlying the \texttt{pan} package is a multivariate extension of regular (univariate) multilevel models; that is, it represents multiple dependent variables simultaneously.
In addition, the model may feature a number of predictor variables with associated fixed and random effects.
Formally, we refer to this model as the multivariate linear mixed-effects model \citep[MLMM; see][]{Schafer2002a}.
The model reads
\begin{equation} 
  \label{eq:pan}
  \mathbf{y}_{ij} = \mathbf{x}_{ij} \boldsymbol\beta + \mathbf{z}_{ij} \mathbf{b}_j + \mathbf{e}_{ij} \text{,}
\end{equation}
\noindent where $\mathbf{y}_{ij}$ is the $(1 \times r)$ vector of responses for person $i$ in group $j$, $\mathbf{x}_{ij}$ and $\mathbf{z}_{ij}$ are $(1 \times p)$ and $(1 \times q)$ vectors of covariate values, $\boldsymbol\beta$ is a $(p \times r)$ matrix of fixed effects, $\mathbf{b}_j$ is a $(q \times r )$ matrix of random effects, and $\mathbf{e}_{ij}$ is a $(1 \times r)$ vector of residuals.
In most cases, the matrix $\mathbf{z}_{ij}$ contains a subset of the values in $\mathbf{x}_{ij}$, and both will contain at least a ``one'' for the regression intercept.
The random effects matrix $\mathbf{b}_{j}$, with columns stacked upon another, is assumed to follow a normal distribution with mean zero and covariance matrix $\boldsymbol\Psi$, independently and identically for all groups.
The vector of residuals $\mathbf{e}_{ij}$ is assumed to follow a normal distribution with mean zero and covariance matrix $\boldsymbol\Sigma$, independently and identically for all individuals.
The MLMM imputes all variables on the left-hand side of the model equation given the variables on the right-hand side (with fixed and random effects).
Only the variables on the left-hand side (i.e., in $\mathbf{y}_{ij}$) may contain missing values, whereas the variables on the right-hand side must be completely observed (i.e., in $\mathbf{x}_{ij}$ and $\mathbf{z}_{ij}$).
In the following, we will distinguish between two broad approaches to MI of incomplete multilevel data using \texttt{pan}'s MLMM (see Table \ref{tab:models} for an illustration).

\input{./tables/notationTable_modelComp}

\subsubsection{Multivariate empty model}
In the first approach, the emphasis is placed on the left-hand side of the model (i.e., the $\mathbf{y}_{ij}$), whereas the right-hand side includes only the intercept ($\mathbf{x}_{ij}=\mathbf{z}_{ij}=1$).
For all variables included on the left-hand side, the MLMM decomposes their variances and covariances into separate between- ($\boldsymbol\Psi$) and within-group portions ($\boldsymbol\Sigma$).
We refer to this approach as the multivariate \emph{empty} model.
This model can be understood as a multivariate variant of the regular empty multilevel model---also known as the null model or the intercept-only model---, in which the dependent variable is also decomposed into between- and within-group components, but the predictor side of the model remains empty.
The upper half of Table \ref{tab:models} contains an example with three variables, each of which may or may not contain missing data.
As can be seen in Table \ref{tab:models}, the three variables decompose into a fixed term common to all persons and groups, a random intercept unique to each group, and an error term unique to each person.
The covariance matrices of random effects and errors, $\boldsymbol\Psi$ and $\boldsymbol\Sigma$, contain the variances of the $\mathbf{y}_{ij}$ and allow for relations between the dependent variables at the group and the person level, respectively.
For that reason, the empty model is especially useful if researchers are interested in estimating relationships at the individual and the group level as in random-intercept models with group-level predictors (see Example 1).

\subsubsection{Full mixed-effects model}
The second approach utilizes both sides of the model.
For all variables included on the right-hand side (i.e., in $\mathbf{x}_{ij}$ and $\mathbf{z}_{ij}$), the MLMM estimates fixed and/or random effects, respectively.
The lower half of Table \ref{tab:models} contains an example with two dependent variables with missing data and one fully observed predictor variable $x_1$ in $\mathbf{x}_{ij}$ and $\mathbf{z}_{ij}$.
As can be seen, the model includes both fixed and random effects for the intercept and $x_1$.
We refer to this model as the \emph{full} mixed-effects model because it includes both random intercepts and slopes where possible.
Note that $x_1$ is not decomposed in this model, and the fixed and random effects represent the overall effects of that variable on the dependent variables.
In order to include separate within- and between-group effects of $x_1$, the variable must be decomposed into between- and within-group portions prior to performing MI (e.g., by including the group mean as an additional predictor).
The full mixed-effects model is particularly useful if the model of interest includes random slopes because the slope variance is represented in the imputation model (see Example 2).

\subsection{Software Alternatives}

A number of software packages have introduced procedures for MI of multilevel data.
The software M\emph{plus} \citep{Muthen2012} implements a two-level model similar to the empty model in \texttt{pan} (denoted H1) as well as a second procedure (denoted H0) for more complex models \citep[e.g., random-slope models; see][]{Asparouhov2010b}.
Joint modeling approaches are also available in SAS \citep{Mistler2013a}, \texttt{REALCOM} \citep{Carpenter2011}, and the R package \texttt{jomo} \citep{Quartagno2016}.
A fully conditional specification of MI is available in the R package \texttt{mice} \citep{vanBuuren2011}.
For some of these packages, it is possible to follow similar analysis steps as outlined in this article for the \texttt{pan} package.
We return to this possibility in a later section.

\section{Example Applications with Multilevel Missing Data}

In order to demonstrate the application of \texttt{pan} for imputing incomplete multilevel data, we made use of the \texttt{mitml} package.
This package provides a more convenient interface for the \texttt{pan} algorithm and some additional tools for handling multiply imputed data sets and combining their results \citep{mitml032}.
Following the imputation, we used the package \texttt{lme4} for estimating the two models of interest \citep{Bates2014}.
We repeated the imputation and estimation in both examples using the popular software M\emph{plus}\footnote{ \setstretch{1.25}
For Example 1, we used H1 imputation, which is equivalent to the multivariate empty model.
For Example 2, we used H0 imputation because a model that was equivalent to the full mixed-effects model could not be specified using H1 imputation.
}.
The results were mostly consistent with those of \texttt{pan} and will not be discussed in detail.
Input files for M\emph{plus} are provided in Supplement A in the supplemental online materials.

The example data set is structured as follows. The first variable (\texttt{ID}) denotes the class membership of each student.
The remaining variables are as described above and may contain different amounts of missing data, which are denoted as \texttt{NA}.

\begin{lstlisting}[style=rcode]
ID MathAchiev ReadAchiev CognAbility SES MathDis ReadDis SchClimate
 1     517.92     547.65          52  70     1.6     1.8       1.25
 1     524.78     633.82          46  40     3.0     2.4       2.50
 1     544.50     474.04          59  34      NA     2.4       1.00
\end{lstlisting}

\noindent
Treating and analyzing multilevel missing data usually involves the following steps.
First, an appropriate imputation model must be specified.
As outlined above, the analysis model must be considered at that point so that the relevant variables, parameters, and auxiliary variables are included in the imputation model.
Second, the imputation procedure must be carried out, resulting in a number of imputed data sets.
Third, the data sets must be analyzed separately, and the resulting parameter estimates are combined according to the rules described in \citeauthor{Rubin1987} (\citeyear{Rubin1987}; for alternatives, see \citealp{Reiter2007a,Carpenter2013}).
These steps can be carried out using the \texttt{mitml} package.
In order to illustrate the impact of different approaches for handling incomplete multilevel data, we also provide the results obtained from single-level MI, which ignores the multilevel structure, and from listwise deletion (LD; i.e., complete case analysis).
The computer code and output files are provided in Supplement B in the supplemental online materials.

\subsection{Example 1: Random-Intercept Model}

In the first example, the model of interest examined the between- and within-group effects of disciplinary problems in math classes (DPM) on math achievement (MA), while controlling for SES at the individual and class level.
\begin{equation} 
  \tag{\ref{eq:ex1}, revisited}
  \mathit{MA}_{ij} = \beta_{0} + \beta_{1} (\mathit{DPM}_{ij} - \overline{\mathit{DPM}}_{j}) + \beta_{2} (\mathit{SES}_{ij}- \overline{\mathit{SES}_{j}}) + \beta_{3} \overline{\mathit{DPM}}_{j} + \beta_{4} \overline{\mathit{SES}}_{j} + \upsilon_{0j} + \epsilon_{ij}
\end{equation}
\noindent
Choosing an appropriate imputation model is straightforward in this case because the multivariate empty model is suitable for random-intercept models in general.
In addition, the empty model includes between- as well as within-group relations as required by the model of interest (in $\boldsymbol\Psi$ and $\boldsymbol\Sigma$; see Table 3).
Recall that the empty model is specified by writing all variables on the left-hand side of the model equation.
In R, the imputation model for this example is set up as follows.

\begin{lstlisting}[style=rcode]
# SETUP: imputation model (variance decomposition model)
fml <- MathAchiev + MathDis + SES + ReadAchiev + CognAbility + ReadDis + 
       SchClimate ~ 1 + (1|ID)
\end{lstlisting}

\noindent
The \texttt{mitml} package uses formula objects to represent the imputation model.
The ``\texttt{$\sim$}'' symbol separates the left- and right-hand side of the model.
The left-hand side contains the three variables of interest and the auxiliary variables (i.e., reading achievement, cognitive ability, ratings of disciplinary problems in reading classes and school climate).
On the right-hand side, the intercept is specified both as a fixed (\texttt{1}) and a random effect (\texttt{1|ID}), where the ``\texttt{|}'' symbol denotes clustering.

For running the \texttt{pan} algorithm, the \texttt{mitml} package offers the function \texttt{panImpute} as its main interface.
The \texttt{pan} algorithm uses Markov chain Monte Carlo (MCMC) techniques to draw replacements for the missing values.
At each iteration of the procedure, a new set of parameters and replacements is simulated.
The distribution from which the replacements are drawn is called the posterior predictive distribution of the missing data \citep{Gelman2013}.
The full procedure is divided into a burn-in phase and an imputation phase \citep[see][]{Enders2010}.
During burn-in, the algorithm performs a number of iterations without saving any imputations, thus ensuring that the parameters of the imputation model have converged to stationary distributions.
In other words, the burn-in phase must be long enough for the algorithm to ``stabilize'' before any replacements are drawn.
Then, during the imputation phase, a number ($m$) of imputed data sets are drawn, each spread a number of iterations apart.
The fact that imputations are not drawn directly from consecutive iterations ensures that the imputed data sets constitute independent random draws from the posterior predictive distribution.
Specifically, consecutive iterations in MCMC are often correlated to some degree (autocorrelation), whereas multiply imputed data sets must be drawn independently of one another.
Thus, the number of iterations chosen between imputations must be large enough for autocorrelation to vanish.

In the first example, we ran \texttt{pan} for 50,000 burn-in iterations, after which $m=100$ imputed data sets were drawn, each spread 5,000 iterations apart.
While these numbers may seem large, recent studies have advocated generating such large numbers of imputations, particularly when large portions of the data are missing \citep{Graham2007,Bodner2008}.
The number of iterations for burn-in and between imputations was chosen such that convergence could be ensured, as described below.
The respective command using \texttt{mitml} was as follows.

\begin{lstlisting}[style=rcode]
# IMPUTATION: 
imp <- panImpute(dat, formula=fml, n.burn=50000, n.iter=5000, m=100, seed=1234)
\end{lstlisting}

\noindent
The \texttt{mitml} package saves the imputation in a special format that is designed to handle large data sets.
In order to obtain a list containing all the imputed data sets, the function \texttt{mitmlComplete} is used.
The necessary command is printed below.

\begin{lstlisting}[style=rcode]
# list of imputed data sets
impList <- mitmlComplete(imp, print="all")
\end{lstlisting}

\subsubsection{Convergence diagnostics}
For the analysis to yield reliable results, it must be ensured that the \texttt{pan} algorithm has converged and that the imputed data sets are approximately independent draws from the posterior predictive distribution \citep[for a detailed discussion of convergence assessment in MCMC, see][]{Gill2014,Jackman2009,Cowles1996}.
The \texttt{mitml} package offers two ways of doing so.
The first option is to examine the potential scale reduction factor \citep[also called $\hat{R}$;][]{Gelman1992} for the parameters of the imputation model.
Originally intended for analyzing multiple MCMC chains, $\hat{R}$ is calculated here by discarding the burn-in iterations and dividing the single MCMC chain for each parameter into multiple segments \citep[see][]{Asparouhov2010a}.
The $\hat{R}$ statistic then compares the variance within and between segments in order to detect a potential ``drifting'' of the chain, that is, chains that are more variable overall than one would expect, based on the variability within segments.
In the \texttt{mitml} package, $\hat{R}$ is included in the summary of an imputed data object.

\begin{lstlisting}[style=rcode]
# DIAGNOSTIC: summary and potential scale reduction
summary(imp)
\end{lstlisting}

\noindent
In addition to the potential scale reduction, the output of \texttt{summary} includes details about the imputation procedure and the missing data rate per variable.
In this example, the output was as follows (truncated for better readability).

\begin{lstlisting}[style=rcode]
Call:

panImpute(data = dat, formula = fml, n.burn = 50000, n.iter = 5000, 
              m = 100, seed = 1234)

Cluster variable:         ID 
Target variables:         MathAchiev MathDis SES ReadAchiev CognAbility ReadDis SchClimate 
Fixed effect predictors:  (Intercept) 
Random effect predictors: (Intercept) 

Performed 50000 burn-in iterations, and generated 100 imputed data sets,
each 5000 iterations apart. 

Potential scale reduction (Rhat, imputation phase):
 
         Min   25%  Mean Median   75%   Max
Beta:  1.000 1.000 1.000  1.000 1.000 1.000
Psi:   1.000 1.000 1.001  1.000 1.001 1.011
Sigma: 1.000 1.000 1.000  1.000 1.000 1.001

Largest potential scale reduction:
Beta: [1,6], Psi: [1,1], Sigma: [1,1]

Missing data per variable:
    ID MathAchiev MathDis SES  ReadAchiev CognAbility ReadDis SchClimate
MD% 0  19.4       61.4    35.0 0          0           21.5    21.7      
\end{lstlisting}

\noindent Ideally, $\hat{R}$ should be close to one for all parameters \citep{Gelman1992}.
If larger values occur (say, above 1.050), a longer burn-in period may be required.
Due to the potentially large number of statistical parameters, the \texttt{mitml} package displays only summary statistics for these values while emphasizing the parameters with the largest $\hat{R}$.
As shown in the output, the $\hat{R}$ was well below $1.050$ for all parameters.
The parameter with the largest $\hat{R}$ was the first diagonal entry of the random-effects covariance matrix $\boldsymbol\Psi$, that is, the intercept variance for math achievement scores ($\hat{R}=1.011$).
However, $\hat{R}$ has been criticized, and large values of $\hat{R}$ need not always indicate poor convergence \citep[e.g.,][]{Geyer1992}.
Therefore, as a second option, diagnostic plots should be considered.
For each parameter in the imputation model, the \texttt{plot} function may produce a trace plot for all iterations during and/or after burn-in, an autocorrelation plot for all iterations after burn-in, and a summary of the parameter's posterior distribution.
The trace plot is a graphical representation of the MCMC chain for each parameter, and it shows the values of that parameter at each iteration.
The autocorrelation plot shows the degree to which consecutive elements of the MCMC chain are correlated (when spread a number of iterations apart).
The posterior summary includes a density plot of the MCMC chain and a number of summary statistics relating to both the MCMC chain and its autocorrelation.
The diagnostic plots can be requested as follows. \\[0ex]

\begin{lstlisting}[style=rcode]
plot(imp, trace="all")
\end{lstlisting}

\begin{figure}[th]
  \label{fig:ex1diag}
  \centering
  \fitfigure{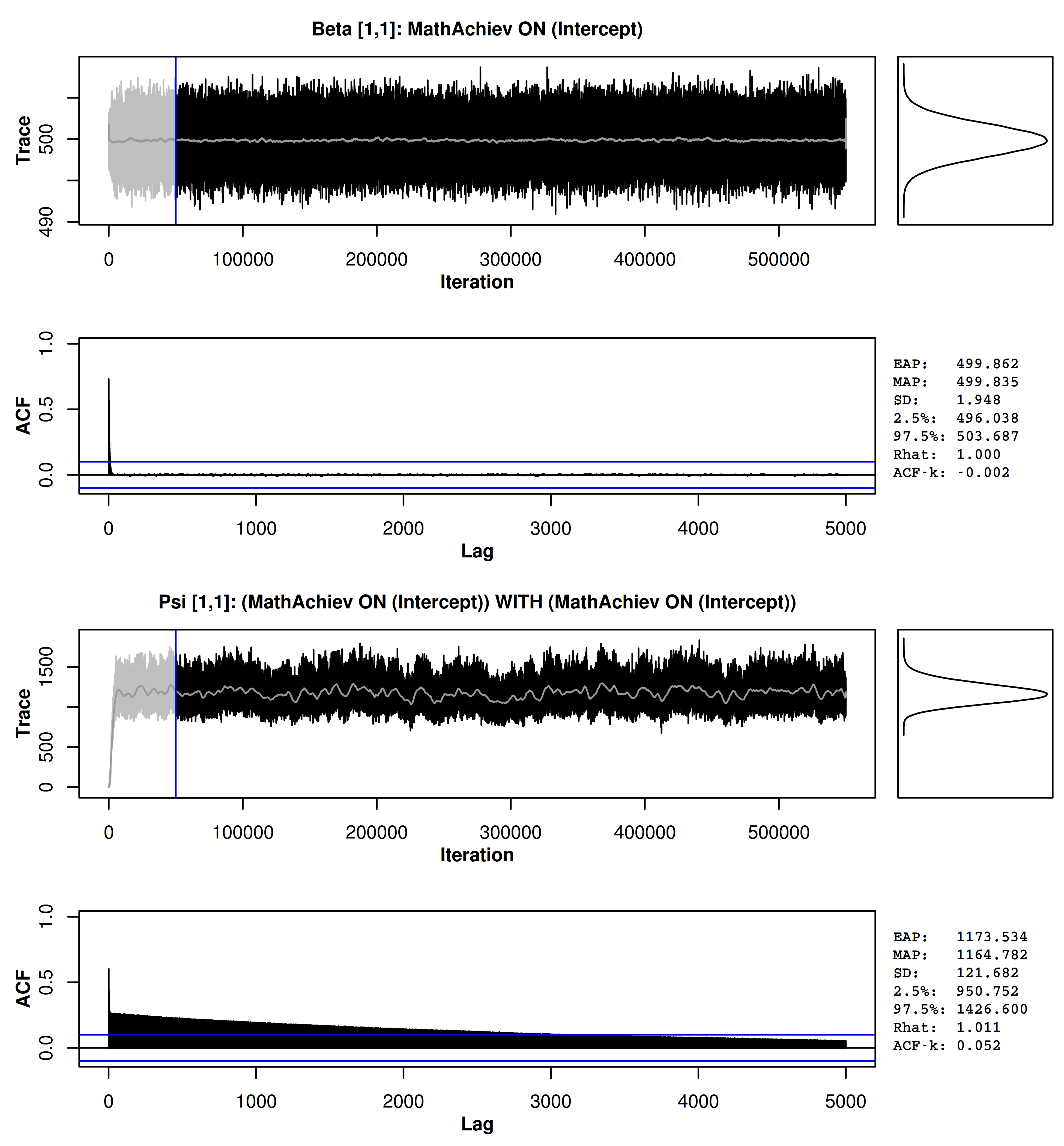}
  \caption{Diagnostic plots for the fixed intercept (top) and the intercept variance (bottom) of math achievement in the imputation model. The trace plot includes all iterations from the burn-in and the imputation phase. The autocorrelation plot and the posterior summaries are calculated only from the imputation phase.}
\end{figure}

\noindent
Here, we discuss the diagnostic plots only for the fixed intercept and the intercept variance for math achievement, which exhibited the worst convergence behavior of all parameters (see Figure~1).
The trace plots showed no sign of ``drifting'' or substantial change after the burn-in phase, indicating that 50,000 iterations were sufficient for the parameters to reach their respective target distributions.
Autocorrelation was quite persistent for the intercept variance but had essentially died out by lag 5,000.
Therefore, imputations spread 5,000 iterations apart could be considered independent.
We concluded that the parameters had converged and that the imputed data sets constituted independent draws from the posterior predictive distribution of the missing data.

\begin{table}[tbp]
  \begin{threeparttable}
    \caption{Estimates of the Intraclass Correlation for the Variables of Interest in Example 1}
    \label{tab:ex1iccs}
    \renewcommand{\arraystretch}{1.4}
    \setlength{\tabcolsep}{9.6pt}
    \newcolumntype{d}{D{.}{.}{3}}
    \small
    \begin{tabular}{lddd} \toprule
    & \multicolumn{1}{c}{Multilevel MI} & \multicolumn{1}{c}{Single-level MI} & \multicolumn{1}{c}{Listwise Deletion} \\ \midrule
    $\mathrm{ICC}_\mathit{MA}$&0.121&0.111&0.115 \\
    $\mathrm{ICC}_\mathit{SES}$&0.122&0.072&0.134 \\
    $\mathrm{ICC}_\mathit{DPM}$&0.179&0.100&0.169 \\ \bottomrule
    \end{tabular}
    \begin{tablenotes}[para,flushleft] ~\\[-1.2ex]
      {\small \textit{Note.} MA = math achievement; SES = socioeconomic status; DPM = disciplinary problems in math classes; ICC = intraclass correlation.
      }
    \end{tablenotes}
  \end{threeparttable}
\end{table}

\begin{table}[tbp]
  \begin{threeparttable}
    \caption{Results from Multilevel MI, Single-Level MI, and Listwise Deletion for Example 1 (Random- Intercept Model)}
    \label{tab:ex1results}
    \renewcommand{\arraystretch}{1.4}
    \setlength{\tabcolsep}{9.0pt}
    \newcolumntype{d}{D{.}{.}{3}}
    \small
    \begin{tabular}{ldddddddd} \toprule
    & \multicolumn{3}{c}{Multilevel MI} & \multicolumn{3}{c}{Single-level MI} & \multicolumn{2}{c}{Listwise deletion} \\ \cmidrule(lr){2-4} \cmidrule(lr){5-7} \cmidrule(lr){8-9}
    &\multicolumn{1}{c}{Estimate}&\multicolumn{1}{c}{SE}&\multicolumn{1}{c}{FMI}
    &\multicolumn{1}{c}{Estimate}&\multicolumn{1}{c}{SE}&\multicolumn{1}{c}{FMI}
    &\multicolumn{1}{c}{Estimate}&\multicolumn{1}{c}{SE} \\ \midrule
    Intercept&502.498&19.254&0.287&502.268&22.326&0.231&505.911&20.063 \\
    $\mathit{SES}_{ij}$&1.065&0.084&0.436&1.054&0.080&0.401&0.849&0.138 \\
    $\overline{\mathit{SES}}_{j}$&2.150&0.264&0.316&2.474&0.303&0.237&1.753&0.275 \\
    $\mathit{DPM}_{ij}$&-20.736&2.032&0.518&-21.609&1.816&0.440&-21.874&3.054 \\
    $\overline{\mathit{DPM}}_{j}$&-41.131&5.035&0.271&-47.165&5.764&0.187&-31.552&5.689 \\ \midrule
    $\textit{Var}(\upsilon_{0j})$&655.957&&&592.132&&&731.860& \\
    $\textit{Var}(\epsilon_{ij})$&8318.936&&&8387.913&&&8299.000& \\ \bottomrule
    \end{tabular}
    \begin{tablenotes}[para,flushleft] ~\\[-1.2ex]
      {\small \textit{Note.} Estimates were significant at $p<.001$; SE = standard error; MA = math achievement; SES = socioeconomic status; DPM = disciplinary problems in math classes; $\upsilon_{0j}$ = random intercepts; $\epsilon_{ij}$ = residuals at Level~1.
      }
    \end{tablenotes}
  \end{threeparttable}
\end{table}

\subsubsection{Intraclass correlations}
Usually the first step in analyzing multilevel data is to estimate the intraclass correlation (ICC) of the variables of interest.
Therefore, before proceeding with the model of interest, we will illustrate the analysis of multiply imputed data sets by fitting intercept-only models for math achievement, SES, and DPM to estimate their ICCs.
In order to obtain final parameter estimates from multiply imputed data sets, the analysis model must be fit separately to each data set, and the resulting estimates must be combined.
In the \texttt{mitml} package, the list of imputed data sets (here \texttt{impList}) can be analyzed by using the functions \texttt{with} and \texttt{within}.
The \texttt{within} function is used to transform the imputed data sets and carry out smaller computations prior to fitting the analysis model.
The \texttt{with} function returns the model fit itself.
The intercept-only model for math achievement can be fit as shown below (for DPM and SES, see Supplement B).
We used the \texttt{lmer} function from the \texttt{lme4} package to fit the analysis models.

\begin{lstlisting}[style=rcode]
# FIT: null model for math achievement
fit <- with( impList, lmer(MathAchiev ~ 1 + (1|ID)) )
\end{lstlisting}

\noindent
This results in a list of 100 fitted analysis models, one for each imputed data set.
The parameter estimates of the fitted models can be combined by using the rules described in \citet{Rubin1987}.
The \texttt{mitml} package implements Rubin's rules in the \texttt{testEstimates} function, which returns the combined estimates for all fixed effects and, when used with \texttt{lme4}, the variance components and the residual ICC (see Supplement B).
The final estimates can be requested as given below.

\begin{lstlisting}[style=rcode]
# final parameter estimates (Rubin's rules)
testEstimates(fit, var.comp=TRUE)
\end{lstlisting}

\noindent
The resulting estimates of the ICCs are presented in Table \ref{tab:ex1iccs} along with the estimates from single-level MI and LD.
Most notably, multilevel MI (using \texttt{pan}) led to much larger estimates of the ICCs than single-level MI, especially for variables with large amounts of missing data (DPM and SES).
This illustrates the importance of accounting for the multilevel structure when conducting MI for multilevel data.
The estimates obtained from LD were closer to those of multilevel MI without any obvious pattern emerging.
These results are consistent with previous research that was based on simulation studies \citep[e.g.,][]{vanBuuren2011a,Taljaard2008}.

\subsubsection{Model of interest}
The procedures outlined above can also be used for fitting the model of interest (Equation \ref{eq:ex1}).
Prior to fitting the model, the group means for DPM and SES must be calculated in each imputed data set, and the student-level variables must be centered around their respective group means.
Such computations can be carried out using \texttt{within} as shown below.

\begin{lstlisting}[style=rcode]
# TRANSFORM: class means for MathDis and SES
impList <- within( impList, { MathDis.CLS <- clusterMeans(MathDis,ID)
                              SES.CLS <- clusterMeans(SES,ID) } )

# TRANSFORM: center student-level predictors
impList <- within( impList, { MathDis.STU <- MathDis - MathDis.CLS
                              SES.STU <- SES - SES.CLS } )
\end{lstlisting}

\noindent
This results in a list of 100 imputed data sets, similar to the original list, but with the group means and the group-mean-centered variables added to each data set.
Finally, the model of interest was fit as shown below using the \texttt{lme4} package (using \texttt{with}).

\begin{lstlisting}[style=rcode]
# FIT: model of interest
fit <- with( impList, lmer(MathAchiev ~ 1 + SES.STU + SES.CLS + MathDis.STU +
                      MathDis.CLS + (1|ID)) )
\end{lstlisting}

\noindent
As before, \texttt{testEstimates} returned the final parameter estimates and inferences.

\begin{lstlisting}[style=rcode]
# final parameter estimates (Rubin's rules)
testEstimates(fit, var.comp=TRUE)
\end{lstlisting}

\noindent
The output of \texttt{testEstimates} includes the final parameter estimates, the MI standard errors, the degrees of freedom and value of the reference $t$ distribution\footnote{\setstretch{1.25}
By default, \texttt{testEstimates} uses the standard $t$ distribution proposed by \citet{Rubin1987}, which provides a test statistic that is appropriate in larger samples.
Alternatively, the degrees of freedom may be adjusted for smaller samples as described in the package documentation \citep[see also][]{Barnard1999,Reiter2007}.
}, the fraction of missing information (FMI), and the relative increase in variance due to nonresponse (RIV).
Even though the FMI is not frequently reported in empirical studies, it holds great value for the interpretation of results and has been recommended as a diagnostic tool for analyzing multiply imputed data sets \citep{Bodner2008}.
The FMI represents the amount of information about an estimand that is lost due to missing data \citep{Allison2001,Enders2010}.
In other words, the FMI shows the loss of ``efficiency'' when estimating parameters from multiply imputed data sets \citep{Savalei2012}.
Similar to the FMI, the RIV denotes the increase in sampling variability in each estimand that can be attributed to missing data \citep[see][]{Enders2010}.
The output for the model of interest is given below.

\begin{lstlisting}[style=rcode]
Call:

testEstimates(model = fit, var.comp = TRUE)

Final parameter estimates and inferences obtained from 100 imputed data sets.

             Estimate Std.Error   t.value        df   p.value       RIV       FMI 
(Intercept)   502.498    19.254    26.098  1210.447     0.000     0.401     0.287 
SES.STU         1.065     0.084    12.614   526.055     0.000     0.766     0.436 
SES.CLS         2.150     0.267     8.065  1429.302     0.000     0.357     0.264 
MathDis.STU   -20.736     2.032   -10.203   372.305     0.000     1.065     0.518 
MathDis.CLS   -41.131     5.035    -8.169  1358.967     0.000     0.370     0.271 

                        Estimate 
Intercept~~Intercept|ID  655.957 
Residual~~Residual      8318.936 
ICC|ID                     0.073 

Unadjusted hypothesis test as appropriate in larger samples.
\end{lstlisting}

\noindent
The results for multilevel MI, single-level MI, and LD are presented in Table \ref{tab:ex1results}.
In general, a higher SES was associated with higher math achievement scores, whereas test scores tended to be lower if students reported disciplinary problems in class.
The estimates at the class level were roughly twice as large as those at the student level.
Single-level MI led to similar estimates of within-group effects, but the estimates of the between-group effects were consistently larger than those obtained from multilevel MI.
Listwise deletion produced larger standard errors (especially at the student level) and smaller estimates of class-level effects.

Researchers are often interested in estimating \emph{contextual} effects, that is, group-level effects when controlling for effects at the student level.
For example, the contextual effect of SES can be calculated simply by subtracting its within-group coefficient from its between-group coefficient \citep{Kreft1995}.
Effects constrained in such a way can be tested against zero using the \texttt{testConstraints} function as shown below.

\begin{lstlisting}[style=rcode]
# contextual effect via model constraints
testConstraints(fit, "SES.CLS - SES.STU")
\end{lstlisting}

\noindent
Testing constrained parameters is based on the delta method \citep[e.g.,][]{Casella2002,Raykov2004}, and the pooled test for multiply imputed data sets is based on the method by \citet{Li1991}\footnote{\setstretch{1.25}
The method by \citet{Li1991} requires that the FMIs are approximately equal across the parameters being tested \citep[see also][]{Licht2010}.
In the present case, the linear constraint being tested has only one component and fulfills this requirement automatically.
}.
For further details, we refer to the package documentation.
The output for testing the contextual effect of SES is printed below.

\begin{lstlisting}[style=rcode]
Call:

testConstraints(model = fit, constraints = "SES.CLS - SES.STU")

Hypothesis test calculated from 100 imputed data sets. The following
constraints were specified:

 SES.CLS - SES.STU

Combination method: D1 

    F.value      df1      df2  p.value      RIV 
     15.297        1 1292.993    0.000    0.365 

Unadjusted hypothesis test as appropriate in larger samples. 
\end{lstlisting}

\noindent
In this example, the contextual effect of SES was statistically significant at $p<.001$ ($F=15.297$, $\mathit{df}_1=1$, $\mathit{df}_2=1293.0$).
Thus, it appeared that classes with a higher SES tended to have higher math achievement scores, even after controlling for SES at the student-level.

Notice that, throughout this example, we used manifest group means as predictor variables in the multilevel analyses.
This is different from the imputation model, where the group-level portions of variables are represented as latent variables (i.e., random effects).
In general, an imputation model based on latent group means (i.e., random effects) yields similar results as one that is based on manifest means, and both can be considered correct imputation models for multilevel data \citep{Ludtkeinpress, Mistler2015, Carpenter2013}.
However, when estimating the model of interest, the predictors' group means may again be considered as latent, and slightly different results are expected for such an analysis model \citep{Asparouhov2006,Ludtke2008}.
A further discussion can be found in Supplement C in the supplemental online materials along with the M\emph{plus} syntax files for fitting the latent analysis model.
In this example, the two analysis models led to essentially the same conclusions.

\subsection{Example 2: Random-Slope Model}

In the second example, the model of interest examined the effect of student's cognitive ability (CA) and socioeconomic status (SES) on students' math achievement scores (MA).
The effect of SES is assumed to be fixed, whereas the effect of cognitive ability is allowed to vary across groups.
\begin{equation} 
\tag{\ref{eq:ex2}, revisited}
\begin{aligned}
  \mathit{MA}_{ij} &= \beta_{0} + \beta_{1} (\mathit{CA}_{ij} - \overline{\mathit{CA}}_{j}) + \beta_{2} \overline{\mathit{CA}}_{j} + \beta_3 (\mathit{SES}_{ij} - \overline{\mathit{SES}}_{j}) + \beta_{4} \overline{\mathit{SES}}_{j} \\
 &+ \upsilon_{0j} + \upsilon_{1j} (\mathit{CA}_{ij} - \overline{\mathit{CA}}_{j}) + \epsilon_{ij}
\end{aligned}
\end{equation}
\noindent
As discussed before, the imputation model must consider the model of interest.
In this example, the effect of cognitive ability is assumed to vary across groups, which must be reflected in the imputation model.
The full mixed-effects model was used for this task  (see Table \ref{tab:models}).
Furthermore, we calculated the group means and the group-mean-centered cognitive ability scores so that we could use them in the imputation model.
This was achieved using \texttt{within} as shown below.

\begin{lstlisting}[style=rcode]
# TRANSFORM: group mean centering (prior to performing MI)
dat <- within(dat, { CognAbility.CLS <- clusterMeans(CognAbility,ID)
                     CognAbility.STU <- CognAbility - CognAbility.CLS } )
\end{lstlisting}

\noindent
Because cognitive ability scores are available for all students, it can be included on the right-hand side of the imputation model, which also allows the slope variance to be specified.
The imputation model was set up as follows.

\begin{lstlisting}[style=rcode]
# SETUP: imputation model (random effects model)
fml <- MathAchiev + SES + ReadAchiev + MathDis + ReadDis + SchClimate ~ 1 + 
       CognAbility.STU + CognAbility.CLS + (1+CognAbility.STU|ID)

\end{lstlisting}

\noindent
The model includes math achievement, SES, and the auxiliary variables on the left-hand side of the equation.
In order to include the slope variance, cognitive ability is featured on the right-hand side, where \texttt{(1+CognAbility.STU|ID)} allows the intercept and the effect of the group-mean-centered cognitive ability scores to vary across groups.

It is worth noting that the MLMM assumes the same random effects structure for \emph{all} dependent variables in the model.
In other words, the full mixed-effects model includes not only the intercepts and slopes for the regressions of math achievement and SES on cognitive ability but also for the four remaining variables.
Thus, users of \texttt{pan} should be wary of including too many variables if the model contains multiple random effects.
The number of parameters can increase rapidly by adding dependent variables or predictors with random effects to the model, possibly requiring a large number of iterations for the model to converge.

As in the first example, the imputation procedure is started by using \texttt{panImpute} while referring to the data set and the model equation.
In this example, we let \texttt{pan} perform 100,000 burn-in iterations, after which we generated $m=100$ imputed data sets, each spread 20,000 iterations apart.
The respective command was as follows.

\begin{lstlisting}[style=rcode]
# IMPUTATION: 
imp <- panImpute(dat, formula=fml, n.burn=100000, n.iter=20000, m=100, seed=1234)
\end{lstlisting}

\noindent
As before, a list of imputed data sets was extracted using \texttt{mitmlComplete}.
The code is not displayed here because it is identical to the previous example (see Supplement B).

\subsubsection{Convergence diagnostics}
\begin{figure}[th]
  \label{fig:ex2diag}
  \centering
  \fitfigure{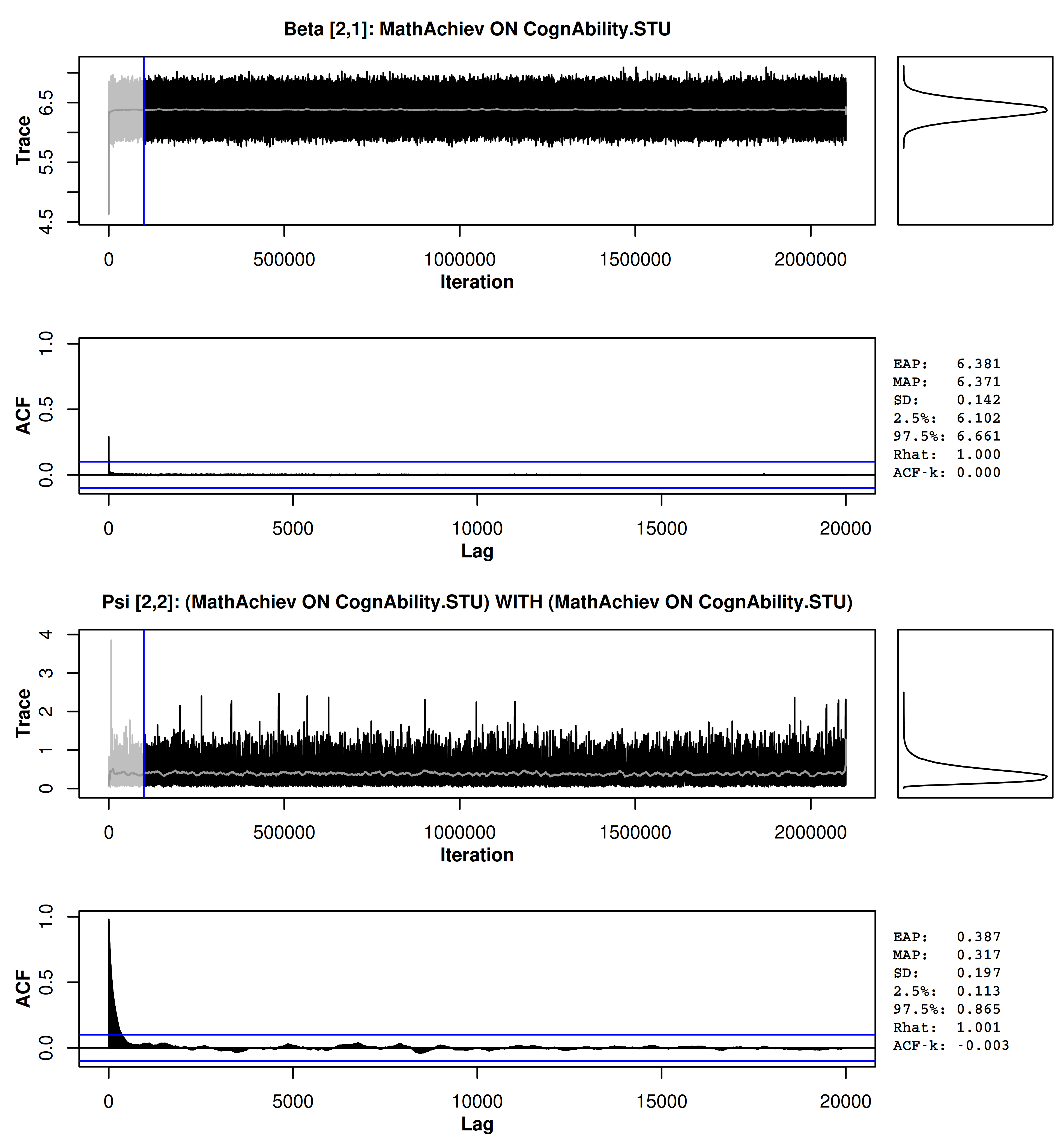}
  \caption{Diagnostic plots for the fixed effect (top) and the slope variance (bottom) for the regression of math achievement on cognitive ability in the full mixed-effects model. The trace plot includes all iterations from the burn-in and the imputation phase. The autocorrelation plot and the posterior summaries are calculated from the imputation phase.}
\end{figure}
Before proceeding with the analysis, it must be ensured that the \texttt{pan} algorithm has converged during burn-in and that the interval between imputations was sufficiently large.
Again, $\hat{R}$ gives an idea of possible problems with convergence and is accessed through the \texttt{summary}.
The largest value of $\hat{R}$ was $1.001$ in this case, indicating that the MCMC chain had become stationary for all parameters.
Examining the diagnostic plots supported this impression but also indicated that some parameters were affected by autocorrelation.
As shown in Figure~2, the parameters related to the variables of interest converged quickly and did not suffer greatly from autocorrelation.
For some parameters, especially the group-level variance components, the autocorrelation was quite persistent but vanished for all parameters with a lag of 15,000 to 20,000 iterations.

\subsubsection{Model of interest}
\begin{table}[tbp]
  \begin{threeparttable}
    \caption{Results from Multilevel MI, Single-Level MI, and Listwise Deletion for Example 2 (Random-Slope Model)}
    \label{tab:ex2results}
    \renewcommand{\arraystretch}{1.4}
    \setlength{\tabcolsep}{8.2pt}
    \newcolumntype{d}{D{.}{.}{3}}
    \small
    \begin{tabular}{ldddddddd} \toprule
    & \multicolumn{3}{c}{Multilevel MI} & \multicolumn{3}{c}{Single-level MI} & \multicolumn{2}{c}{Listwise deletion} \\ \cmidrule(lr){2-4} \cmidrule(lr){5-7} \cmidrule(lr){8-9}
    &\multicolumn{1}{c}{Estimate}&\multicolumn{1}{c}{SE}&\multicolumn{1}{c}{FMI}
    &\multicolumn{1}{c}{Estimate}&\multicolumn{1}{c}{SE}&\multicolumn{1}{c}{FMI}
    &\multicolumn{1}{c}{Estimate}&\multicolumn{1}{c}{SE}\\
    \midrule
    Intercept                    &84.573&21.364&0.108&79.514&20.562&0.113&96.792&25.956 \\
    $\mathit{CA}_{ij}$           &6.114 &0.150 &0.185&6.138 &0.154 &0.232&6.250 &0.185  \\
    $\overline{\mathit{CA}}_{j}$ &7.608 &0.521 &0.164&7.549 &0.501 &0.160&7.714 &0.584  \\
    $\mathit{SES}_{ij}$          &0.605 &0.077 &0.463&0.599 &0.075 &0.445&0.578 &0.079  \\
    $\overline{\mathit{SES}}_{j}$&1.081 &0.261 &0.289&1.254 &0.291 &0.277&0.749 &0.253  \\ \midrule
    $\textit{Var}(\upsilon_{0j})$                &485.050 &&  &417.802 &&         &540.924 & \\
    $\textit{Var}(\upsilon_{1j})$                &1.528   &&  &1.381   &&         &1.572   & \\
    $\textit{Cov}(\upsilon_{0j},\upsilon_{1j})$&0.333   &&  &1.470   &&         &5.535   & \\
    $\textit{Var}(\epsilon_{ij})$                &6452.506&&  &6547.366&&         &6287.600& \\
    \bottomrule
    \end{tabular}
    \begin{tablenotes}[para,flushleft] ~\\[-1.2ex]
      {\small \textit{Note.} Estimates for the fixed effects were significant at $p<.001$; SE = standard error; CA = cognitive ability; SES = socioeconomic status; $\upsilon_{0j}$ = random intercepts; $\upsilon_{1j}$ = random slopes; $\epsilon_{ij}$ = residuals at Level~1.
      }
    \end{tablenotes}
  \end{threeparttable}
\end{table}
In order to estimate the model of interest, the student-level variables were centered around their group means (using \texttt{within}), and the model was fit using the \texttt{lme4} package (using \texttt{with}).
We changed the method for estimating the multilevel model from restricted maximum likelihood (REML) to full information maximum likelihood (FIML) because the model comparison that was conducted as a later step in this analysis required that the analysis models were estimated using FIML.
The code for fitting the model of interest is given below.

\begin{lstlisting}[style=rcode]
# FIT: model of interest
fit <- with( impList, lmer(MathAchiev ~ 1 + CognAbility.STU + CognAbility.CLS +
                      SES.STU + SES.CLS + (1+CognAbility.STU|ID), REML=FALSE) )

\end{lstlisting}

\noindent
The final parameter estimates and inferences were obtained using \texttt{testEstimates}. 
These are presented in Table \ref{tab:ex2results}, along with the estimates from single-level MI and LD.
Students with higher cognitive ability (as compared with their class average) tended to score higher on the math achievement test after controlling for SES.
This relation appeared to vary substantially across groups.
In comparison, single-level MI produced lower estimates of the intercept and slope variance and a slightly larger estimate of the class-level effect of cognitive ability.
For LD, the estimates of the fixed effects and variance components were slightly different from those obtained with MI but comparable altogether.
Results obtained using the H0 imputation in M\emph{plus} yielded results similar to those produced by \texttt{pan}\footnote{ \setstretch{1.25}
These differences were negligible for most parameters, but M\emph{plus} produced a large estimate of the slope variance, $\textit{Var}(\upsilon_{1j})=2.277$.
Despite the large similarities, there are some subtle differences between \texttt{pan} and the H0 imputation in M\emph{plus}.
For example, M\emph{plus} uses ``least informative'' priors for H1 but improper priors for H0, which cannot be specified using \texttt{pan}.
However, preliminary simulations could not replicate any difference between \texttt{pan} and M\emph{plus}.
A more in-depth exploration of these (relatively minor) differences was beyond the scope of this article and will be left for future research.
}.

When estimating multilevel models with random slopes, researcher are often interested in whether or not the regression coefficients vary substantially across groups.
For this purpose, likelihood-ratio tests (LRTs), which compare the model of interest with an alternative model that constrains the slope variance to zero, are often conducted \citep[see][]{Snijders2012a}.
A method for pooling the LRT across multiply imputed data sets was suggested by \citet{Meng1992}.
This procedure is accessible in \texttt{mitml} through the \texttt{testModels} function.
The alternative model is similar to the model of interest, but only the intercept is allowed to vary across groups.
The code for fitting the alternative model is given below.

\begin{lstlisting}[style=rcode]
# FIT: null model without random slopes
fit.null <- with( impList, lmer(MathAchiev ~ 1 + CognAbility.STU + CognAbility.CLS +
                           SES.STU + SES.CLS + (1|ID), REML=FALSE) )

\end{lstlisting}

\noindent
The two models can be compared using \texttt{testModels}, where \texttt{method="D3"} calls the procedure by \citet{Meng1992}.
The respective command was as follows.

\begin{lstlisting}[style=rcode]
# LRT for nonzero slope variance
testModels(fit, fit.null, method="D3")
\end{lstlisting}

\noindent
The output of \texttt{testModels} for testing the slope variance is printed below.

\begin{lstlisting}[style=rcode]
Call:

testModels(model = fit, null.model = fit.null, method = "D3")

Model comparison calculated from 100 imputed data sets.
Combination method: D3 

     F.value       df1       df2   p.value       RIV 
       5.119         2 10386.237     0.006     0.157 
\end{lstlisting}

\noindent The pooled LRT was statistically significant at $p=.006$ ($F=5.119$, $\mathit{df}_1=2$, $\mathit{df}_2=10386.2$) indicating that the slope variance was statistically different from zero.
Thus, it appeared that students with different cognitive ability may differ more or less strongly in their math achievement scores, depending on the class to which they belong.
It may be interesting to examine the determinants of this variation, for example, teachers' attributes or aspects of the learning environment.
However, for the purpose of this article, we will not discuss these questions in detail.
Research has shown that the LRT for variance components may suffer from low statistical power \citep[see][]{Stram1994,LaHuis2009}.
However, there are currently very few options for performing hypothesis tests for variance components with multiply imputed data sets other than Meng and Rubin's \citeyearpar{Meng1992} method.

\section{Analyzing Imputations Generated by Alternative Software}

As outlined above, there are a number of software alternatives for generating imputations for multilevel missing data, some of which are similar in scope to \texttt{pan}, and some of which provide further support for categorical, ordinal or group-level variables.
For example, if the model of interest also includes categorical variables with missing data, researchers may prefer using the R packages \texttt{jomo} or \texttt{mice}, or standalone software such as M\emph{plus}.
In general, the analysis steps presented here can be carried out on multiply imputed data sets irrespective of their origin.
The requirement for using \texttt{mitml}'s analysis functions is that the multiply imputed data sets are represented as a ``list'' of data sets in R.
This can be achieved by either generating imputations using its wrapper functions, or by converting the imputed data into a list of data sets.
The \texttt{mitml} package currently includes wrapper functions for \texttt{pan} (\texttt{panImpute}) and \texttt{jomo} (\texttt{jomoImpute}) as well as functions to convert imputed data sets generated by \texttt{mice} (\texttt{mids2mitml.list}).
For other software packages, however, the conversion must be performed manually (e.g., using \texttt{long2mitml.list}, or \texttt{as.mitml.list}).
The use of these functions is illustrated in the documentation of the package.
In most applications, using the wrapper functions is recommended because it allows for using the tools for convergence diagnostics provided by \texttt{mitml}.

\section{Discussion}

Even though multilevel models are frequently used in psychology and the social sciences, MI of multilevel missing data is seldom discussed in the applied literature.
As a result, listwise deletion, single-level MI, and ad hoc methods for representing the clustered data structure prevail in research practice (e.g., the dummy-indicator approach) even though research has shown that these methods can result in distorted parameter estimates in subsequent multilevel analyses.
In the present article, two empirical examples were used to illustrate the application of the two R packages \texttt{pan} and \texttt{mitml} to multilevel data.
In Example 1, we discussed the application of \texttt{pan} to random-intercept models and for estimating between- and within-group effects.
In Example 2, we focused on MI for multilevel models with random slopes and on estimating and testing the slope variance.
We believe that researchers can benefit greatly from incorporating \texttt{pan} in their statistical analyses.
Specifically, \texttt{pan} allows the special features of multilevel data to be preserved, a practice that is essential for obtaining reliable estimates from multilevel analyses and for understanding their results.
Moreover, \texttt{pan} allows researchers to use all of the available information in the data and to include auxiliary information without altering the model of interest.
By contrast, many interesting features of multilevel models may be distorted or even lost when using simpler methods for handling multilevel missing data.
For example, the results from Example 1 showed that parameter estimates can be distorted if the imputation model ignores the multilevel structure of the data.

The field of statistical software is always in motion, and there continue to be a number of promising developments regarding multilevel MI.
However, some problems still provide challenges for the future.
For example, using multilevel MI can be difficult if missing data occur on predictor variables in models with random slopes or interaction effects.
\citet{Graham2012} recommended that MI for models with random slopes should be conducted separately for each group using single-level MI.
\citet{Schafer2001a} proposed that incomplete predictor variables be treated as outcome variables in the imputation model, thus accepting a (possibly small) bias for the slope variance \citep[see also][]{Grund2016}.
To mitigate this problem, it has been suggested to generate imputations for predictor variables in such a way that they are consistent with the model of interest (\citealp[e.g.,][]{Goldstein2014,Wu2010}; see also \citealp{Bartlett2015}).
These methods may provide an improvement over current implementations of multilevel MI in complex multilevel models with random slopes and missing values in predictor variables \citep{Erlerinpress}.
Unfortunately, they are currently not available in standard software.

Even though many algorithms exist for MI of multilevel data, the analysis often remains a challenge when software does not provide the tools for combining the results from multiply imputed data sets.
Using the \texttt{mitml} package, we provided examples for combining simple parameter estimates, model comparisons, and model constraints with multiply imputed data sets.
In addition to Rubin's rules \citeyearpar{Rubin1987}, the package implements the procedures commonly referred to as $D_1$ \citep{Li1991,Reiter2007}, $D_2$ \citep{Li1991a}, and $D_3$ \citep{Meng1992}, which can be used for testing a variety of statistical hypotheses that potentially involve multiple parameters simultaneously (e.g., model comparisons).
Nonetheless, open questions remain about how some statistical quantities can be estimated from multiply imputed data sets.
For example, it is not yet clear how researchers can obtain measures of the goodness-of-fit of multilevel models, which are often used for model selection (e.g., the model deviance, AIC or BIC).
Such procedures might be based on the methods by \citet{Li1991a} and \citet{Meng1992}, or on variations thereof \citep{Licht2010}, but clear recommendations have not yet been made in the literature \citep[see also][]{Consentino2010,Grund2016a}.

The treatment of multilevel missing data offers many challenges, and state-of-the-art procedures are often not very accessible unless researchers are deeply familiar with missing data and MI.
We hope that the present article will provide guidance for applied researchers and promote the use of modern missing data techniques such as MI. 
In general, we believe that \texttt{pan} is a powerful tool for treating multilevel missing data because many features of typical research questions can easily be represented in \texttt{pan}'s MLMM.
Future research should devote attention to increasing the accessibility of modern methods for handling and analyzing missing data.
Currently, the use of MI in multilevel research, while largely desirable, is often hindered by the lack of accessible software and appropriate tools for analyzing multiply imputed data sets in real-world research scenarios.
For future studies, the topic of multilevel missing data yields many interesting research questions that have yet to be explored.

\bibliographystyle{apacite}
\bibliography{/home/simon/Dokumente/Projekte/zotero_fullbib}

\end{document}

%% file: arXiv_titlePage.tex
This article is published in \emph{SAGE Open}, and the print version is available for open access using the link below. This manuscript is the author's personal manuscript after peer review, which was submitted to the journal upon acceptance. It is not the version of record and may not exactly replicate the print version.

~\\[1ex]

The print version of the article can be found online:

\url{http://sgo.sagepub.com/content/6/4/2158244016668220}

~\\[1ex]

The supplemental materials to the article are available at:

\url{http://sgo.sagepub.com/content/spsgo/suppl/2016/10/25/6.4.2158244016668220.DC1/sgo.pdf}

~\\[1ex]

The correct citation for this article is:

Grund, S., L\"udtke, O., \& Robitzsch, A. (2016). Multiple imputation of multilevel missing data: An introduction to the R package pan. \emph{SAGE Open}, \emph{6}(4), 1--17. doi: 10.1177/2158244016668220

%% file: tables/notationTable_modelComp.tex
\begin{table}[tbp]
  \begin{threeparttable}
    \caption{Two Multivariate Linear Mixed-Effects Models for Missing Data}
    \label{tab:models}
    \renewcommand{\arraystretch}{1.5}
    \setlength{\tabcolsep}{2.0pt}
    \small
    \begin{tabular}{c} \toprule
      \emph{General notation}
      \\

      $\mathbf{y}_{ij} = \mathbf{x}_{ij} \boldsymbol\beta + \mathbf{z}_{ij} \mathbf{b}_j + \mathbf{e}_{ij}$
      \\[1.8ex]
      \midrule

      \emph{Multivariate empty model}
      \\[1ex]

      $\begin{array}{c}
        \begin{bmatrix}
          y_{1} & y_{2}  & y_{3}
        \end{bmatrix}_{ij} \\
        \text{\scriptsize{target variables}}
      \end{array}
      \begin{array}{c}
        = \\ {}
      \end{array}
      \begin{array}{c}
        \begin{bmatrix}
          \beta_1 & \beta_2 & \beta_3
        \end{bmatrix} \\
        \text{\scriptsize{fixed effects (intercepts)}}
      \end{array}
      \begin{array}{c}
        + \\ {}
      \end{array}
      \begin{array}{c}
        \begin{bmatrix}
          b_1 & b_2 & b_3
        \end{bmatrix}_j \\
        \text{\scriptsize{random effects (intercepts)}}
      \end{array}
      \begin{array}{c}
        + \\ {}
      \end{array}
      \;\,
      \begin{array}{c}
        \begin{bmatrix}
          e_{1} & e_{2}  & e_{3}
        \end{bmatrix}_{ij} \\
        \text{\scriptsize{residuals}}
      \end{array}$
      \\[6ex]

      $\begin{bmatrix} b_1 & b_2 & b_3 \end{bmatrix}_j
      \sim N(\mathbf{0},\boldsymbol\Psi)
      \quad \text{and} \quad
      \begin{bmatrix} e_{1} & e_{2} & e_{3} \end{bmatrix}_{ij}
      \sim N(\mathbf{0},\boldsymbol\Sigma)$
      \\[3.5ex]

      \midrule

      \emph{Full mixed-effects model}
      \\[1ex]

      $\begin{array}{c}
        \begin{bmatrix}
          y_{1} & y_{2}
        \end{bmatrix}_{ij} \\
        \text{\scriptsize{target variables}}
      \end{array}
      \begin{array}{c}
        = \\ {}
      \end{array}
      \begin{array}{c}
        \begin{bmatrix}
          1 & x_{1}
        \end{bmatrix}_{ij}
        \begin{bmatrix}
          \beta_{01} & \beta_{02} \\ \beta_{11} & \beta_{12}
        \end{bmatrix} \\
        \text{\scriptsize{fixed effects (intercepts, slopes)}}
      \end{array} \;\;
      \begin{array}{c}
        + \\ {}
      \end{array}
      \begin{array}{c}
        \begin{bmatrix}
          1 & x_{1}
        \end{bmatrix}_{ij}
        \begin{bmatrix}
          b_{01} & b_{02} \\ b_{11} & b_{12}
        \end{bmatrix}_j \\
        \text{\scriptsize{random effects (intercepts, slopes)}}
      \end{array}
      \begin{array}{c}
        + \\ {}
      \end{array} \;\;
      \begin{array}{c}
        \begin{bmatrix}
          e_{1} & e_{2}
        \end{bmatrix}_{ij} \\
        \text{\scriptsize{residuals}}
      \end{array}$
      \\[8ex]

      $\begin{bmatrix} b_{01} & b_{11} & b_{02} & b_{12} \end{bmatrix}_j
      \sim N(\mathbf{0},\boldsymbol\Psi)
      \quad \text{and} \quad
      \begin{bmatrix} e_{1} & e_{2} \end{bmatrix}_{ij}
      \sim N(\mathbf{0},\boldsymbol\Sigma)$
      \\[3.5ex]

      \bottomrule
    \end{tabular}
    \begin{tablenotes}[para,flushleft] ~\\[-1.2ex]
      {\small \textit{Note.} The predictor $x_1$ is assumed to be completely observed. Vectorization of the random-effects matrix $\mathbf{b}_j$ is achieved by stacking its columns.}
    \end{tablenotes}
  \end{threeparttable}
\end{table}